\documentclass[runningheads]{llncs}

\usepackage{times}
\usepackage{cite} 
\usepackage{epsfig}
\usepackage{graphicx}
\usepackage{amssymb}
\usepackage{pifont}
\usepackage{soul} % Add soul package for highlighting
\usepackage{amsmath}
\usepackage{hyperref}
\usepackage{amsmath}
\usepackage{amssymb}
\usepackage{dsfont}
\newcommand{\app}{\raise.17ex\hbox{$\scriptstyle\sim$}}
\usepackage{multirow}
\usepackage{lipsum}
\newcommand\blfootnote[1]{%
  \begingroup
  \renewcommand\thefootnote{}\footnote{#1}%
  \addtocounter{footnote}{-1}%
  \endgroup
}
\makeatletter
\def\blfootnote{\gdef\@thefnmark{}\@footnotetext}
\makeatother

\usepackage{booktabs} % For better looking tables
\begin{document}
\title{HoloHisto: End-to-end Gigapixel WSI Segmentation with 4K Resolution Sequential Tokenization}
% \author{Submission 1515}
\author{
Yucheng Tang\inst{1*} \and
Yufan He\inst{1} \and
Vishwesh Nath\inst{1} \and
Pengfeig Guo\inst{1} \and
Ruining Deng\inst{2} \and
Tianyuan Yao\inst{2} \and
Quan Liu\inst{2} \and
Can Cui\inst{2} \and
Mengmeng Yin\inst{3} \and
Ziyue Xu\inst{1} \and
Holger Roth\inst{1} \and
Daguang Xu\inst{1} \and
Haichun Yang\inst{3} \and
Yuankai Huo\inst{2,3}
}
%
% \authorrunning{Y. Tang et al.}
\institute{Anonymous}

\institute{Nvidia \and
Vanderbilt University \and
Vanderbilt University Medical Center}
\maketitle              % typeset the header of the contribution
\begin{abstract}

In digital pathology, the traditional method for deep learning-based image segmentation typically involves a two-stage process: initially segmenting high-resolution whole slide images (WSI) into smaller patches (e.g., $256\times256$, $512\times512$, $1024\times1024$) and subsequently reconstructing them to their original scale. This method often struggles to capture the complex details and vast scope of WSIs. In this paper, we propose the holistic histopathology (HoloHisto) segmentation method to achieve end-to-end segmentation on gigapixel WSIs, whose maximum resolution is above 80,000$\times$70,000 pixels. HoloHisto fundamentally shifts the paradigm of WSI segmentation to an end-to-end learning fashion with 1) a large (4K) resolution base patch for elevated visual information inclusion and efficient processing, and 2) a novel sequential tokenization mechanism to properly model the contextual relationships and efficiently model the rich information from the 4K input. To our best knowledge, HoloHisto presents the first holistic approach for gigapixel resolution WSI segmentation, supporting direct I/O of complete WSI and their corresponding gigapixel masks. Under the HoloHisto platform, we unveil a random 4K sampler that transcends ultra-high resolution, delivering 31 and 10 times more pixels than standard 2D and 3D patches, respectively, for advancing computational capabilities. To facilitate efficient 4K resolution dense prediction, we leverage sequential tokenization, utilizing a pre-trained image tokenizer to group image features into a discrete token grid. To assess the performance, our team curated a new kidney pathology image segmentation (KPIs) dataset with WSI-level glomeruli segmentation from whole mouse kidneys. From the results, HoloHisto-4K delivers remarkable performance gains over previous state-of-the-art models.

% considerably below the gigapixel viewing scale. This approach, while effective for localized analysis, falls short when confronted with the intricate details and expansive scale of whole slide images (WSI). Presently, there is no end-to-end framework tailored for the complete segmentation of WSIs. To address the gap, we introduce HoloHisto, advocating the first holistic approach to histopathology segmentation. HoloHisto redefines WSI segmentation by supporting direct I/O of complete WSI and their corresponding gigapixel masks. Under HoloHisto platform, we unveil random 4K sampler that transcends ultra-high resolution, delivering 31/10 times more pixels than standard 2D/3D patches for advancing computational capabilities. To facilitate efficient 4K resolution dense prediction, we leverages sequential tokenization, utilizing a pre-trained image tokenizer to group image features into a discrete token grid. As such, HoloHisto-4K delivers remarkable performance gains over previous state-of-the-art models. Furthermore, we provide an extensively pathologists annotated dataset. The HoloHisto platform, along with the code and dataset, will be made available for open-source.

% \blfootnote{* equal contribution}
%Source code and the segmentation model will be available at:. 

\end{abstract}

\begin{figure}[t]
\centering
\includegraphics[width=\textwidth]{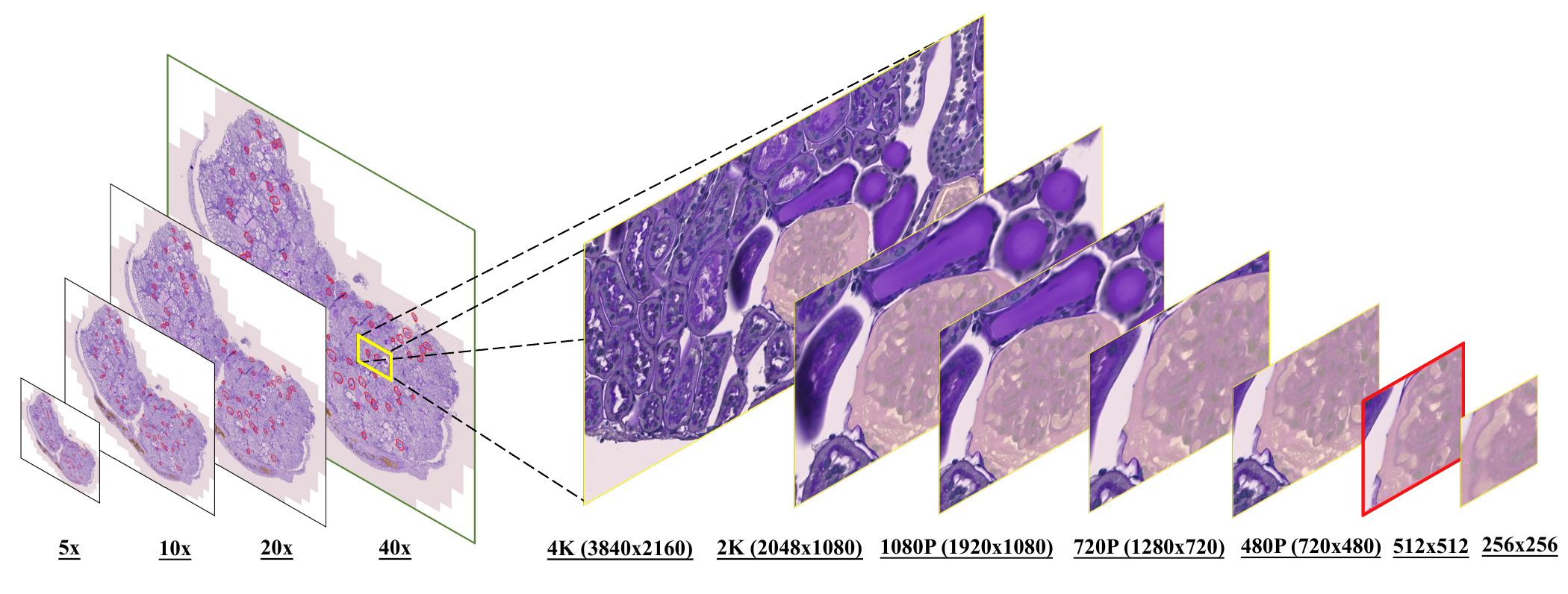}

\caption{The field of view provided by different resolutions, such as the conventional  $512\times512$ patch (indicated by a red box) and $256\times256$, reveals only a limited range of details within the target tissue structures. In contrast, ultra-high resolution (4K) images offer a more comprehensive view of interrelations and can serve as a foundational visual constitute for WSI analysis.}

\label{fig:fig1}
\end{figure}

\section{Introduction}
\label{Introduction}
Digital pathology, a rapidly evolving field of medical vision research, has seen a transformative advancement with large vision models (LVMs)~\cite{wang2023review,ma2024segment,bai2023sequential}. This significant advent created new demands for high-quality perception, which is crucial for microscopic (e.g., whole slide) image computing. However, current models are limited to the capability of dissecting and interpreting small pre-defined patches within images~\cite{deng2023omni,chen2022self}. Typically, pre-processed tiles are confined to dimensions of $512\times512$ pixels or resampled to smaller dimensions of $224\times224$ defined by some predominating frameworks~\cite{dosovitskiy2020image,oquab2023dinov2,caron2021emerging}, which restrict the scopes of tissue details that can be captured. The absence of rich information hinders the model's performance, particularly impacting tasks of detecting small objects and dense prediction~\cite{deng2023omni,feng2023artificial}. For instance, the detection and segmentation of complete medullas under kidney WSI will degrade by more than 10\% in DSC by using a height and width of 512 or completely fail without patching pre-defined ROI~\cite{deng2024prpseg}. This scalability, especially when dealing with gigapixel whole slide image (WSI), remains a bottleneck in comprehensive and efficient computing analysis. To date, there are no established gold standard datasets for segmenting gigapixel WSIs, resulting in a lack of comprehensive end-to-end methods in this histopathology research. 

To include more information, a higher resolution is necessary as shown in Fig.~\ref{fig:fig1}. Nevertheless, modeling ultra-high definition (UHD) images (e.g. beyond 4K resolution) is extremely challenging~\cite{chen2021multimodal,esser2021taming,shen2022high}. High-resolution dense prediction requires a balance of strong context information extraction and model efficiency~\cite{cai2023efficientvit}. The computational cost of convolutional and transformer models, despite their significant benefits, has quadratic increases in the demand for computational resources~\cite{tang2022self}. This presents a critical scaling challenge for processing whole slide images. Therefore, high-quality image dense prediction requires models capable of understanding both global composition and locality of interactions at a compression rate. 

In this work, we propose the HoloHisto framework, debuting the holistic approach to redesign histopathology image segmentation with three key features:

\noindent\textbf{The Holistic Approach:}
We developed an end-to-end workflow for training and inferencing gigascale WSI, introducing a novel learning paradigm to the field of WSI analysis.. HoloHisto is designed to handle inputs and outputs of any size, regardless of whether they are (WSIs) or smaller patches. By leveraging cuCIM, our dataloader facilitates real-time reading of WSIs at various magnification levels and supports random foreground patching, tiling, or augmentation, enhancing the flexibility and efficiency of our approach. Our approach is capable of dynamically creating datasets online from one or multiple WSIs, potentially comprising an unlimited number of images during training. This approach does not depend on pre-defined cropping strategies, offering a more flexible and scalable solution for training models on large-scale datasets. In the inference stage, it can generate the corresponding gigapixel output.

\noindent\textbf{Architecture: }
We design an efficient backbone tailored for segmenting UHD images. First, we employ the sequential tokenizer for learning discrete visual tokens from perceptually rich constituents, streamlining towards 4K resolution dense prediction. Second, to model the long discrete tokens from these UHD images, we propose to use a two-stage ViT architecture that incorporates multi-scale attention~\cite{cai2023efficientvit}, which uses ReLU linear attention instead of the inefficient Softmax attention.

\noindent\textbf{Data:}
As a significant effort to improve gigapixel WSI computing, our pathologists addressed the critical gap in the availability of imaging data. We present Kidney Pathology Image Segmentation (KPIS), the dataset that facilitates the diagnosis and treatment of chronic kidney disease (CKD). Annotations are performed at the WSI level, serving as a foundation benchmark for developing cutting-edge image segmentation technologies.  

In summary, this paper explores a new learning paradigm of WSI segmentation: 1) HoloHisto framework that is capable of paralleling tile processing with direct WSI I/O, 2) scalable segmentation backbone with sequential tokenizer for ultra-high resolution images, and 3) gigapixel WSI annotation dataset as a foundational benchmark.

\section{Related Works}
\textbf{Pathology Segmentation.} 
 Recent advances in deep learning with CNN and transformers~\cite{hara2022evaluating,gadermayr2017cnn} achieve significant improvement in the field of pathology segmentation. Several works were proposed to address the challenges of microscopic imaging data, including H\&E stained pathology images, fluoresce data, or other cell imaging modalities~\cite{bueno2020glomerulosclerosis,israel2023foundation}. Numerous datasets for cell and tissue segmentation, including MoNuSeg~\cite{kumar2019multi} and NEPTUNE~\cite{barisoni2013digital}, are available for identifying a variety of glomerular structures. In addition, instance segmentation is developed in the general cell imaging domain~\cite{ma2023multi}. However, most current approaches focus on analyzing local tiles at a uniform magnification level, including nuclei, glomeruli, or tubules. This results in a notable gap in the segmentation of disease-related regions across entire whole slide images. Despite limited exploration or established efficacy in the field, we introduce a segmentation dataset and methodology designed for comprehensive WSI segmentation.
 
\noindent\textbf{Foundation Vision Models.}
Inspired by the achievement of large language models (LLMs)~\cite{esser2021taming,van2017neural}, many endeavors~\cite{bai2023sequential,guo2024data} have been made to develop foundation vision models. With the development of transformer or state space models~\cite{gu2023mamba}, sequence modeling became the de facto way for modeling visual sentences~\cite{bai2023sequential}, which enabled the uniform modeling of various vision tasks. In this work, we explore large vision models (LVM) for digital pathology with two key features: (1) a pre-trained vector quantized generative adversarial networks (VQGAN)~\cite{esser2021taming} that enables scalable tokenization for the ultra-high-resolution image at a compression rate; (2) an efficient multi-scale attention module for long sequence representation learning. 

\begin{figure}[t]
\centering
\includegraphics[width=\textwidth]{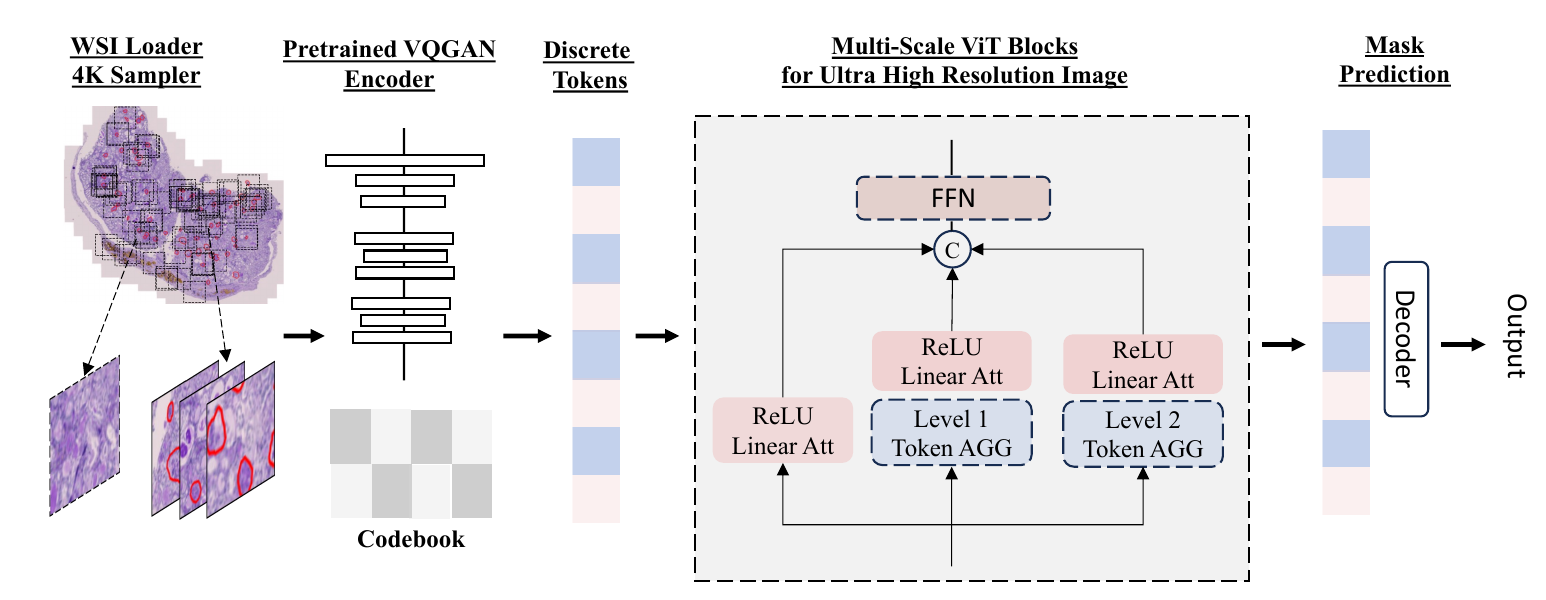}

\caption{\textbf{HoloHisto-4K backbone}. To enable scalable encoding of ultra-high-resolution images (4K), our approach uses a pre-trained convolutional VQGAN to learn the from visual parts into discrete tokens, this design follows autoregressive LVM that enables compression while retaining high perception quality. We employed multi-scale linear attention as an efficient way to capture long discrete visual tokens for high-resolution dense prediction. }

\label{fig:fig2}
\end{figure}

\section{Approach}
In this work, we propose a holistic framework for segmenting gigapixel WSI. In addition, to model ultra-high resolution representation for dense prediction, we propose a model architecture for high-quality perception learning: 1) use the sequence tokenization for learning 4K visual parts at compression scale; 2) train ViT blocks with linear multi-scale attention.  We summarize our approach in Fig~\ref{fig:fig2}.

\subsection{Sequence Tokenization}
To enable scalable modeling of ultra-high-resolution images while circumventing the quadratic increase in complexity associated with the scan-line order of patches, a discrete approach is essential. This method should efficiently encode visual elements and enable the sampling of high-quality perceptual representations from a probabilistic distribution. 
Inspired by neural discrete representation learning~\cite{van2017neural} and Vector Quantised (VQGAN)~\cite{esser2021taming}, we employ an image tokenizer. This tokenizer maps input images to a semantic, discrete token codebook through a quantization layer. This technique captures the rich constituents of images, effectively reducing the expression length of original 4K resolution images. Consequently, it enables efficient modeling of global interrelations. 

Let the given UHD input be denoted by \(x\), which exists in the space \(\mathbb{R}^{H' \times W' \times 3}\). This image can be decomposed into a grid of codebook vectors \(z_{\text{enc}}\), within the domain \(\mathbb{R}^{h' \times w' \times d_z}\). Here, \(d_z\) represents the number of dimensions for each code. We approximate a given image \(x\) by \(\hat{x} = G(z_q)\).

To obtain \(z_q\), we start with the encoding \(\hat{z} = E(x)\), which resides in the space \(\mathbb{R}^{h' \times w' \times d_z}\). Following this, we apply an element-wise quantization \(q(\cdot)\) to each spatial code \(\hat{z}_{ij}\) within \(\mathbb{R}^{n_z}\), aligning it with its nearest entry \(z_k\). The process is formulated as:
\begin{equation}
z_q = q(\hat{z}) := \underset{z_k \in Z}{\arg\min} \|\hat{z}_{ij} - z_k\|
\end{equation}
where \(z_q \in \mathbb{R}^{h' \times w' \times d_z}\) 

% In our experiments given 4K ($3840 \times 2160$) images, we have two token number configurations: 14,400, and 3,600 for compression factors of 24 and 48, respectively. 
\subsection{Linear Multi-Scale Attention}
High-resolution dense prediction models require strong representation learning capability with good efficiency. Instead of widely used Softmax attention~\cite{dosovitskiy2020image}, ReLU attention\cite{katharopoulos2020transformers} provides linear complexity, which offers the flexibility of multi-scale modules for high-resolution dense prediction. Following the efficientViT~\cite{cai2023efficientvit} design, we make transformer blocks consisting of 2-stage multi-scale ReLU attention and FNN layers. The 3 hierarchical multi-scale ReLU attention can be expressed as:
\begin{equation}
A_i = \frac{\text{ReLU}(Q_i) \left(\sum_{j=1}^{N} \text{ReLU}(K_i)^T V_j\right)}{\text{ReLU}(Q_i) \left(\sum_{j=1}^{N} \text{ReLU}(K_j)^T\right)}
\end{equation}

The calculations for the terms \(\left(\sum_{j=1}^{N} \max(0, K_j)^T V_j\right)\) and \(\left(\sum_{j=1}^{N} \max(0, K_j)^T\right)\) need to be performed only once.

\subsection{HoloHisto: End-to-end framework}
\begin{figure}[t]
\centering
\includegraphics[width=\textwidth]{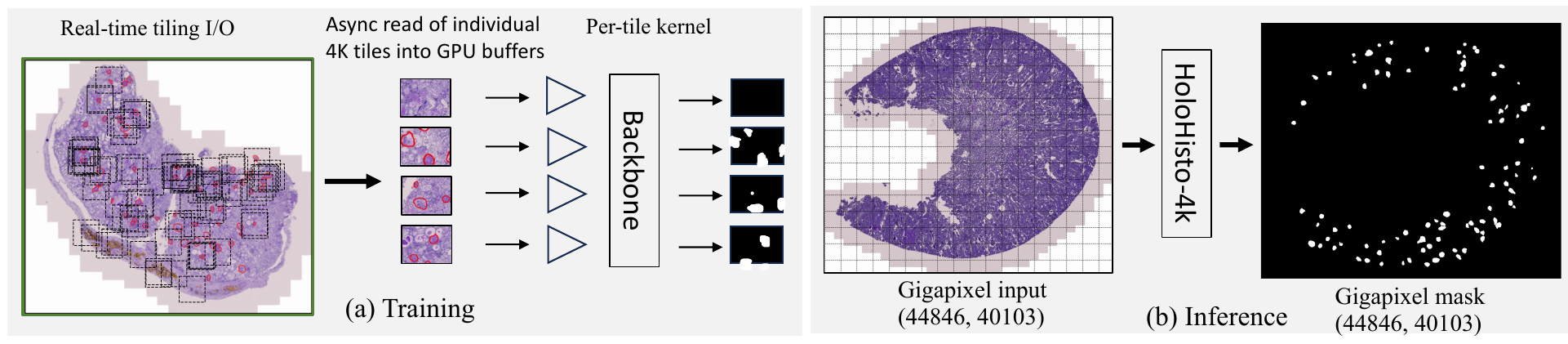}
\caption{The holistic approach for gigapixel WSI segmentation. HoloHisto takes the entire kidney slide image as input, it supports real-time reading of multi-magnification levels and tile sampling. In the inference stage, HoloHisto presents a global gigapixel segmentation mask as output.}
\label{fig:fig3}
\end{figure}

 The complete pipeline of training and inference is demonstrated in Fig~\ref{fig:fig3}.
 
\noindent\textbf{Training Paradigm.}
In prior studies~\cite{chen2022scaling,deng2023omni}, pathology image training has been performed using pre-cropped patches of a fixed size over selected regions of interest (ROIs). This offline preprocessing, used before the training and inference phases for gigapixel images, results in the model repeatedly learning from the same patches in each epoch.
In this work, we introduce a random sampling paradigm for digital pathology image loader based on the cuCIM\footnote{\url{https://developer.nvidia.com/cucim-for-image-io-processing}} multidimensional processing unit. During the training phase, a foreground mask is created using a thresholding technique. Subsequently, we randomly extract ROIs at a 4K resolution from the identified foreground areas. The dataloader then compiles a dataset from one or several whole slide images (WSIs). As the number of training epochs increases, the framework is capable of sampling a virtually "unlimited" number of patches from the WSIs.

\noindent\textbf{Inference with WSI.}
During the inference stage, HoloHisto is capable of processing the entire WSI.  The dataloader seamlessly reads the designated magnification level and isolates the foreground regions through thresholding. Subsequently, HoloHisto performs the foreground tiling with or without overlap, and loads individual tiles into one-dimensional GPU buffers, then positions them correctly within a pre-allocated GPU array until predictions have been made for all tiles. Finally, the predicted masks for each 4K tile can be allocated back onto the WSI space.

\section{Experiments and Results}
\subsection{Datasets}
\label{dataset}
\textbf{Kidney Pathology Image Segmentation (KPIs).} The KPIs challenge cohort includes 60 high-resolution WSIs of whole mouse kidneys derived from 20 rodents, including three CKD disease models in addition to normal kidneys. Tissues were stained with Periodic Acid-Schiff (PAS) and captured at 40$\times$ magnification using the Leica SCN400 Slide Scanner. This diverse cohort allows for comprehensive analysis across different CKD models and normal conditions, providing a rich dataset for advancing research in renal pathology image segmentation. These WSIs are derived from four different groups of mouse models, each representing a different condition or stage of CKD. More information about KPIS studies and annotations is in the supplementary material.

\noindent\textbf{NEPTUNE~\cite{barisoni2013digital}.} The public dataset consists of 1751 Region of Interest (ROI) images that are extracted from 459 Whole Slide Images (WSIs) from 125 patients diagnosed with Minimal Change Disease. These images underwent manual segmentation for six structurally normal pathological features. Each image is at $3000\times3000$ resolution.

\begin{table}[t]
\centering
\scriptsize
\setlength{\tabcolsep}{2mm}
\renewcommand\arraystretch{1}
\caption{Segmentation results of the KPIS dataset on testing cases. Except HoloHisto, evaluations of WSI on baseline models are performed using predictions from non-overlapping tiles, which are mapped back to the WSI space. Dice similarity coefficient scores (\%) are reported. The difference between the reference (Ref.) method and benchmarks for WSI is statistically evaluated by Wilcoxon signed-rank test.}
\resizebox{0.9\textwidth}{!}{
\begin{tabular}{l|cccc|ccc}
\hline
\multirow{2}{*}{Method}   &\multirow{2}{*}{56Nx}  &\multirow{2}{*}{DN} &\multirow{2}{*}{NEP25} &\multirow{2}{*}{Normal} &\multicolumn{3}{|c}{All} \\
\cline{6-8}
 & & &  &  &4K Patch-wise &WSI & Statistic.\\
\hline
U-Nets~\cite{isensee2021nnu} &88.29    &87.14  &84.91   &90.12  &87.62 &62.19 & p$<$0.01\\
UNETR~\cite{hatamizadeh2021unetr} &88.96   &87.47  &85.81   &88.46  &87.68 &64.25 & p$<$0.01\\
SegFormer~\cite{xie2021segformer}  &88.58  &89.86   &89.97   &90.33  &89.69 &65.01 & p$<$0.01\\
 DeepLabV3~\cite{deeplabv3plus2018} &88.41  &89.84  &89.90  &89.91  &89.52 &68.22 & p$<$0.01\\
SwinUNETR-V2~\cite{he2023swinunetr} &89.04    &91.02   &89.05     &90.08   &89.80 &71.58 & p$<$0.01\\
SAM-ViT-B~\cite{kirillov2023segment} &89.98  &90.19   &90.13   &90.23 &90.13 &75.19 & p$<$0.01\\
SAM-ViT-H~\cite{kirillov2023segment} &90.93 &91.04 &90.18 &90.05 &90.55  &77.24 & p$<$0.01\\
\hline
HoloHisto-4K &\textbf{93.77}  &\textbf{92.45} &\textbf{94.81}  &\textbf{94.12} &\textbf{93.79} &\textbf{84.54} &\textbf{Ref.} \\
\hline
\end{tabular}
}
\label{tab:tab1}
\end{table}

\begin{table}[t]
\centering
\scriptsize
\setlength{\tabcolsep}{2mm}
\renewcommand\arraystretch{1}
\caption{NEPTUNE segmentation performance. HoloHisto experiments are conducted under $3000\times3000$ according to the dataset's original resolution. Dice similarity coefficient scores (\%) are reported.}
\resizebox{0.9\textwidth}{!}{

\begin{tabular}{l|cccccc|c}
\hline
\multirow{1}{*}{Metrics}   &\multicolumn{1}{c}{DT}  &\multicolumn{1}{c}{PT} &\multicolumn{1}{c}{CAP} &\multicolumn{1}{c}{TUFT} &\multicolumn{1}{c}{VES} &\multicolumn{1}{c|}{PTC} &\multicolumn{1}{c}{Average} \\
\hline
SAM ViT-B Binary~\cite{kirillov2023segment}			    &79.82 & 85.58    &93.58 &93.09    &83.29 &74.90 &  86.09   \\
UNet multi-scale~\cite{Ronneberger_2015_UNet}		&81.11 &89.85    &96.70 &96.66    &85.03 &77.19  & 87.76    \\
DeepLabV3~\cite{deeplabv3plus2018}			            &81.05 &89.90    &96.77 &96.69    &85.35 &78.04  & 87.97  \\
SwinUNETR-V2~\cite{he2023swinunetr}			    &81.10 &89.02    &96.74 &85.29    &85.33 &78.68  & 86.03  \\
SAM ViT-B multi-scale~\cite{kirillov2023segment}           &81.38 &89.01    &96.90 &96.79    &85.25 &78.58  & 87.99  \\
SAM ViT-H multi-scale~\cite{kirillov2023segment}			&81.40 &90.58    &97.00 &96.95    &85.91 &79.05  & 88.48\\
\hline
HoloHisto-4K			&\textbf{82.14} &\textbf{90.88}    &\textbf{97.06} &\textbf{96.99}    &\textbf{86.11} &\textbf{79.93} &\textbf{88.85}  \\
\hline
\end{tabular}
}

\label{tab:tab2}
\end{table}

\subsection{Experiments}
We conduct a comparative study of the proposed HoloHisto in two datasets: KPIS and the publicly available tissue segmentation NEPTUNE~\cite{barisoni2013digital}. For the evaluation of KPIS dataset, we present comparisons among conventional tile-based segmentation frameworks and calculated metrics in each 4K patch. In addition, to show the effectiveness of the WSI-level segmentation, we compute the Dice score on the entire WSI foreground. We re-trained baseline models including U-Nets~\cite{Ronneberger_2015_UNet}, UNETR~\cite{hatamizadeh2021unetr}, swinunetr-V2~\cite{he2023swinunetr},SegFormer~\cite{xie2021segformer}, and SAM variants~\cite{kirillov2023segment}. We choose these methods based on groups of CNN, transformer and foundational backbones.
\subsection{Ultra-high Resolution Analysis}

\noindent{\bf KPIS. }
Table~\ref{tab:tab1} shows the quantitative result for the binary segmentation task of high-resolution images in the KPIS dataset. We compared HoloHisto to various baselines including CNN, and Transformer-based methods. We evaluated the metrics in two formats: 1) calculate Dice scores under the 4K resolution patch, HoloHisto is trained and inferenced in 4K patch, baseline methods are performed in $1024\times1024$, which is the best scale for SAM and others. HoloHisto consistently outperforms state-of-the-art pathology segmentation backbone. Along with the ablation study on resolution, we observe the higher resolution patch dimensions, the larger margin is obtained from HoloHisto, indicating the effectiveness of the high-quality perception modeling brought by the tokenizer and efficientViT. In Table~\ref{tab:tab3}, we show the ablative experiments result of components design of sequence tokenizer and ReLU linear attention compared to linear projection and multi-head self-attention (MHSA) in vanilla ViT.

\noindent{\bf NEPTUNE.}
We conducted additional experiments on the existing public dataset NETUNE. Among 6 different scales of tissues, HoloHisto surpasses baseline models consistently. HoloHisto experiments are performed in $3000\times3000$ at its largest resolution from the data source, baselines used $1024\times1024$ sliding window inference from their best training strategy. The Dice scores are reported in Table~\ref{tab:tab2}.

\subsection {\bf End-to-end Prediction}
% We compared difference experiment settings for loading dataset: 1)conventional pre-processing for WSI and fetching tiles without overlap. 

\begin{figure}[t]
\centering
\includegraphics[width=0.95\textwidth]{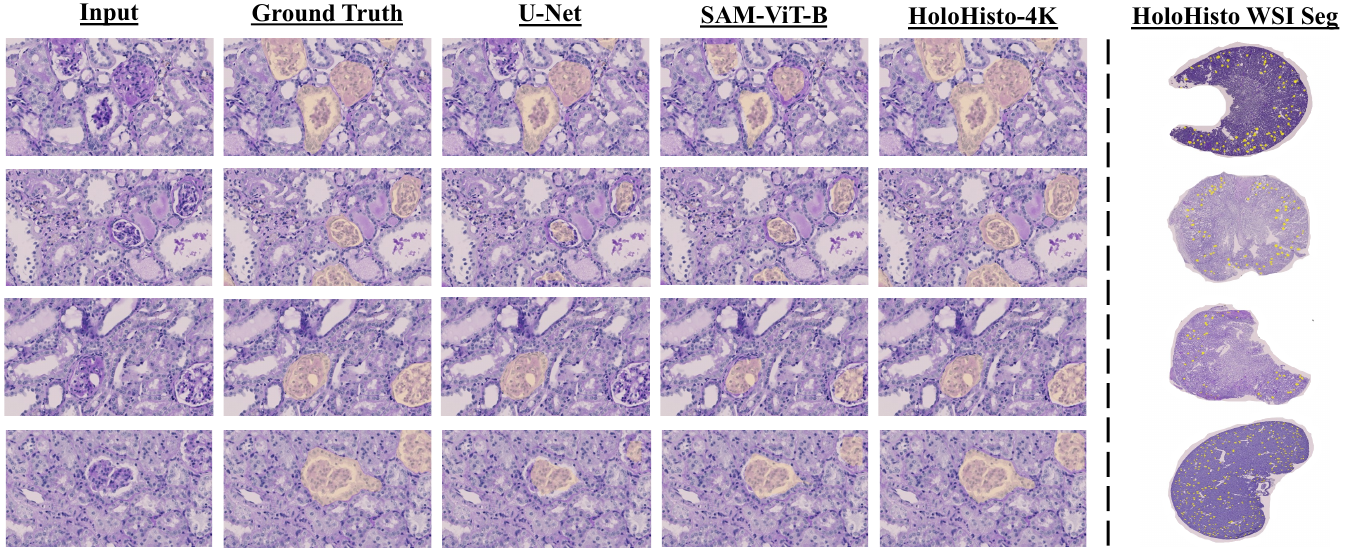}
\caption{Kidney pathology segmentation qualitative results. We show the comparisons of three different approaches from CNN, ViT-based SAM and our HoloHisto. The right column shows HoloHisto's capability of outputting the entire WSI segmentation.}

\label{fig:fig4}
\end{figure}

\noindent{\bf Comparison with pre-tiling}
In Table~\ref{tab:tab3}, the WSI handling section shows the results of using a 4K random sampler with cuCIM and MONAI dataloader versus the tiling strategy, we observe a larger margin of improvement using the end-to-end framework. The visualization of complete WSI is shown in Fig.~\ref{fig:fig4} right panel.

\begin{table}[t]
\centering
\scriptsize
\setlength{\tabcolsep}{2mm} 
\renewcommand\arraystretch{1}
\caption{Ablation Study of approach modules. Dice score of 4k tiles and WSIs are reported for ultra-high resolution encoder and WSI handling configurations, respectively.}
\resizebox{1.0\textwidth}{!}{%
\begin{tabular}{ccccccccccc}
% \cline{1-5} \cline{7-11}
\multicolumn{5}{c}{\textbf{Ultra-high Resolution Encoder}} & \multicolumn{1}{c}{} & \multicolumn{5}{c}{\textbf{WSI Handling}} \\
\cline{1-2} \cline{4-5} \cline{7-8}\cline{10-11}
\multicolumn{2}{c}{Tokenizer} & \multicolumn{1}{c}{} &\multicolumn{2}{c}{ReLU Att} & \multicolumn{1}{c}{} &\multicolumn{2}{c}{Sampler} & \multicolumn{1}{c}{} & \multicolumn{2}{c}{Resolution} \\
\cline{1-5} \cline{7-11}
Linear Proj+MHSA  &90.45 & &Linear Proj+ReLU Att &91.85 & & tile sampler &80.87 & & 2K &79.41 \\
\cline{1-5} \cline{7-11}
Tokenizer+MHSA &92.94 & & Tokenizer+ReLU Att &93.79 & & rand sampler &84.54 & & 4K &84.54 \\

\cline{1-5} \cline{7-11}
\end{tabular}
}
\label{tab:tab3}
\end{table}

% \begin{table}[t]
% \centering
% \caption{Ablation Study.}
% \label{tab:tab1}
% \begin{tabular}{@{}cccccccc@{}}
% \toprule
% \multicolumn{4}{c}{Ultra-high Resolution Encoder}              & \multicolumn{4}{c}{WSI Handling}                    \\ \midrule
% \multicolumn{2}{c}{Tokenizer}          & \multicolumn{2}{c}{Multi-scale ViT} & \multicolumn{2}{c}{Sampler}       & \multicolumn{2}{c}{Resolution} \\
% \midrule
% \multicolumn{2}{c}{w/ Tokenizer~\cite{zhou2021nnformer}} & \multicolumn{2}{c}{w/ multi-scale ViT} & \multicolumn{2}{c}{w/ random sampler} & \multicolumn{2}{c}{4K UHD} \\
% \multicolumn{2}{c}{0.9094}             & \multicolumn{2}{c}{0.7418}            & \multicolumn{2}{c}{0.0000}          & \multicolumn{2}{c}{0.0000}      \\
% \bottomrule
% \end{tabular}
% \end{table}

\section{Discussion and Conclusion}
This work tackles the fundamental task of segmenting histopathology images, a task that formerly relied on complex pipelines and was restricted to the analysis of small patches. We propose a holistic approach to segment gigapixel images with direct WSI I/O. To model the ultra-high resolution images within loaded WSI, we propose to use a sequential tokenizer, which encodes images as a composition of perception parts and thereby avoids the quadratically increased complexity. In addition, we evaluate the linear ReLU multi-scale attention instead of the Softmax attention for 4K UHD image tokens. In experiments, we exhibit the first WSI-level segmentation via a 4K image patch sampler and show the effectiveness and capability of HoloHisto-4K by outperforming state-of-the-art approaches. Towards the development of cutting-edge computational research, we also provide the gold-standard pathologist annotated dataset as a WSI segmentation benchmark.

\textbf{Limitation.} It is important to note that we employed the natural image pre-trained sequence tokenizer, where the learned codebook is not for histopathology images. It is still rather challenging to achieve pathology LVM, limiting the model performance and application to WSI analysis. Therefore, we will continue to explore generalist models for pathology vision tasks.


% \noindent {\bf Acknowledgements.}
% sec \\\\\\ 

% \bibliographystyle{unsrt}
\bibliographystyle{splncs04}
\bibliography{paper}

\end{document}